\documentclass[12pt]{amsart}
\usepackage{amsbsy,amssymb,amscd,amsfonts,latexsym,amstext,delarray,
amsmath,graphicx} 
\input xypic

\usepackage{color}

\numberwithin{equation}{section}

\def\bG{{\mathbb G}}

\def\bW{{\mathbb W}}

\def\N{{\mathbb N}}

\def\R{{\mathbb R}}
\def\Z{{\mathbb Z}}

\def\cC{{\mathcal C}}

\def\cE{{\mathcal E}}
\def\cF{{\mathcal F}}
\def\cG{{\mathcal G}}

\def\cI{{\mathcal I}}
\def\cJ{{\mathcal J}}

\def\cN{{\mathcal N}}
\def\cO{{\mathcal O}}
\def\cP{{\mathcal P}}

\def\cT{{\mathcal T}}

\def\fG{{\mathfrak G}}

\title{Categorical Hopfield Networks}
\author{Matilde Marcolli}
\date{2021}

\address{California Institute of Technology, Pasadena \\ USA}
\email{matilde@caltech.edu}

\begin{document}
\maketitle

\begin{abstract}
This paper discusses a simple and explicit toy-model example of the categorical Hopfield
equations introduced in previous work of Manin and the author. These describe 
dynamical assignments of resources to networks, where resources are 
objects in unital symmetric monoidal categories and assignments are realized by summing
functors. The special case discussed here is based on computational resources (computational
models of neurons) as objects in a category of deep neural networks (DNN), with a simple choice of the 
endofunctors defining the Hopfield equations that reproduce the usual updating of the
weights in DNNs by gradient descent. 
\end{abstract}


\section{Introduction: categorical Hopfield equations}\label{IntroSec}

In \cite{ManMar} a categorical model of neuronal networks with assigned resources was developed
based on Segal's formalism \cite{Segal} of categories of summing functors and Gamma-spaces, see also \cite{Mar21}, \cite{MarTsao18}.
Resources are modeled according to \cite{CoFrSp16}, as categories $\cC$ endowed with either a unital
symmetric monoidal structure, or in a more restrictive case categories with a coproduct (categorical
sum) and zero-object (both initial and terminal). Given a finite set $X$, summing functors $\Phi$ are 
assignments of resources, that is objects in $\cC$, to subsets of $X$ that are additive (in the coproduct 
or the monoidal operation of $\cC$) over disjoint subsets. 

\smallskip

In this model, the category $\Sigma_\cC(X)$ of summing functors and invertible natural transformations
introduced in \cite{Segal} 
is viewed as the parameterizing space of all possible such assignments of $\cC$-resources up to
equivalence. 
Instead of considering just a finite set $X$, one can consider a network (finite directed graph) $G$
and summing functors $\Phi$ that assign $\cC$-resources to subnetworks $G'\subset G$, with
a corresponding category $\Sigma_\cC(G)$ of network summing functors. The category
$\Sigma_\cC(G)$ (or a suitable subcategory representing additional constraints on the
summing functors such as conservation laws at vertices, inclusion-exclusion, or other compositional 
properties) is considered as the ``configuration space" of the model.

\smallskip

A dynamical system is then introduced on this configuration space, in the form of an equation
modeled on the Hopfield equations of network dynamics. Here the variable of the equation
is a summing functor (an assignment of resources) and the dynamics is determined by an
initial condition (an initial assignment) and endofunctors of the category of resources that
determine the dynamics and play a role analogous to the matrix of weights in the usual
Hopfield equations (which we review in \S \ref{HopSec} below). 

\smallskip

The setting presented in \cite{ManMar} is quite abstract and involves choices of
a category of resources $\cC$, a suitable subcategory of network summing functors $\Sigma_\cC(G)$,
and a collection of endofunctors of $\cC$ that determine the Hopfield dynamics. We summarize in
\S \ref{ResourceSec} and \S \ref{DynSummingSec} 
the notion of network summing functors and the categorical Hopfield equations introduced 
in \cite{ManMar}. The main idea is to replace the variables of the classical Hopfield equations,
which describe firing activity and firing rates of neurons, with variables that describe assignments
of resources of a certain type to the nodes (and/or edges) of a network, so that the equations
can be used to describe how these assignments evolve in time, under specific transformation
of both the network and the associated resources.  The purpose of
the present paper is to illustrate this mechanism in a more concrete setting, with a specific choice
of a category $\cC$ of computational models of the nodes of the network, motivated by
computational models of the neuron via deep neural networks developed in  \cite{BeSeLo}.

\smallskip

The categorical Hopfield equations we discuss as our main example in this paper should not
be regarded as a realistic model of categorical dynamics for a neuronal network, rather it is
designed to be just a simple toy model where the categorical setting is easy to describe and
properties of the resulting equations can be easily identified. 

\smallskip

The main results presented in this paper consist of: the construction of a category
of resources based on deep neural networks (DNNs); the explicit form of the categorical
Hopfield equations with this category of resources, in the cases with or without
a leak-term in the equations; the analysis of fixed points of these equations and
the identification of the backpropagation mechanism for the weights of DNNs based
on gradient descent as a special case of this categorical Hopfield dynamics. 

\smallskip
\subsection{Resources and summing functors}\label{ResourceSec}

The mathematical theory of resources and resource convertibility developed in
\cite{CoFrSp16} and \cite{Fr17} describes resources as objects in a unital
symmetric monoidal category $\cC$, with the unit representing the empty
resource and the monoidal operation describing the combination of independent
resources. Convertibility between resources $A$ and $B$ is described by the
existence of a nonempty set of morphisms, ${\rm Mor}_\cC(A,B)\neq \emptyset$.
A measure of the convertibility of resources is provided by the preordered abelian
semigroup given by the set of isomorphism classes $[A]$ of objects $A$ in $\cC$
with the operation $[A]+[B]=[A\oplus B]$ with $\oplus$ the monoidal structure of $\cC$,
and with $[A]\succeq [B]$ whenever convertibility from $A$ to $B$ holds, namely
whenever ${\rm Mor}_\cC(A,B)\neq \emptyset$.

\smallskip

The notion of summing functors was introduced by Segal in \cite{Segal} as part of the mechanism
of Gamma-spaces, a homotopy theoretic construction of spectra from categories with
certain properties (sums and zero-object in Segal's original formulation, or unital symmetric
monoidal categories in the more general formulation of Thomason, \cite{Tho82}, \cite{Tho95}).
Summing functors are assignments $\Phi$ of objects in a category $\cC$ of resources to
subsets of a given finite set $X$, that are additive on disjoint subsets (with respect to the
monoidal structure $\oplus$ of $\cC$),
\begin{equation}\label{summing}
 \Phi(X_1 \sqcup X_2)=\Phi(X_1)\oplus \Phi(X_2), 
\end{equation} 
with $\Phi(\emptyset)=0$, that are functorial under inclusions of subsets of $X$. 
Morphisms in the category of summing functors are invertible
natural transformations. Summing functors are completely determined by the
collection of objects $\{ \Phi(x) \}_{x\in X}$ in $\cC$ and invertible natural transformations
of summing functors by isomorphisms $\{ \eta_x \}_{x\in X}$ of these objects. This
identifies the category $\Sigma_\cC(X)$ of $\cC$-valued summing functors on a given finite set $X$
of cardinality $n$ with the $n$-fold cartesian product $\hat\cC^n$ of the category $\hat\cC$ with the
same objects as $\cC$ and the isomorphisms of $\cC$ as morphisms.

\smallskip

This simple notion of summing functors can be generalized in various ways when
the finite set $X$ is replaced by a finite directed graph $G$. The resulting categories
of ``network summing functors" are described in \cite{ManMar}. 

\smallskip

We focus here on one particular case among those described in \cite{ManMar}, namely the
case where the network summing functors are compatible with a compositional structure on
the category of resources described by an operad or a properad. We need to recall the
setting we consider on the category of resources before introducing the appropriate notion
of network summing functors.

\smallskip
\subsubsection{Properads and operads in categories}\label{OpPropSec}

\begin{figure}
 \begin{center}
 \includegraphics[scale=0.45]{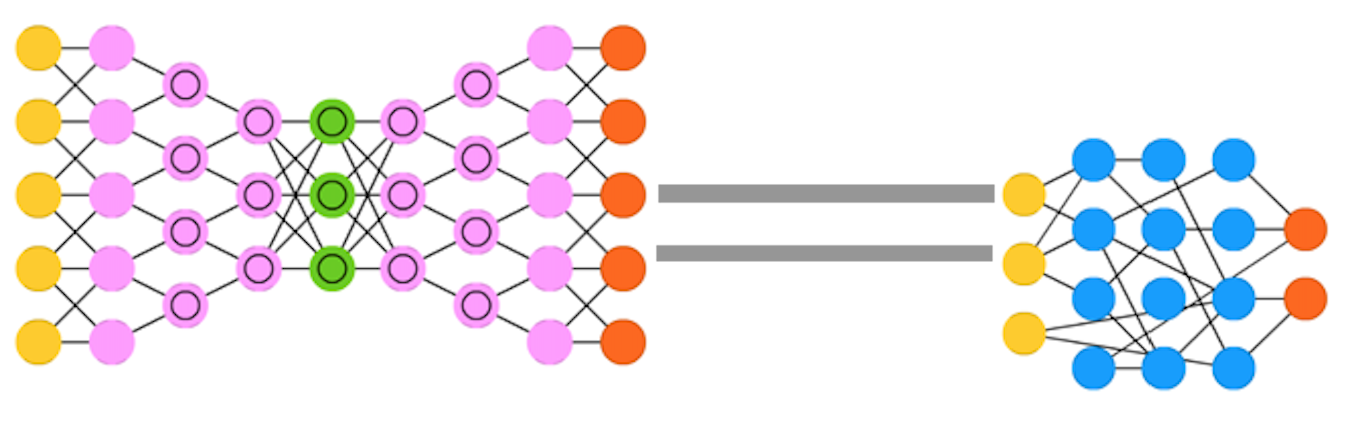}  
 \end{center} 
 \caption{Example of properad composition $\circ^{3,4}_{1,2}: \cP(5,5)\times \cP(3,2)\to \cP(6,5)$ connecting the
 third and fourth outputs of the first object (machine) to the first and second inputs of the second one. The inputs of the resulting object (machine) are the inputs of the first one and the remaining inputs of the second,
 and the resulting outputs are the outputs of the second together with the remaining outputs of the first. \label{ImageTryPropFig}}
 \end{figure}

Let $\text{Cat}$ denote the category of small categories. A properad in $\text{Cat}$  is a collection
$\cP=\{ \cP(n,m) \}_{n,m\in \N}$ of small categories with composition functors
\begin{equation}\label{properad}
 \circ^{i_1,\ldots, i_\ell}_{j_1,\ldots, j_\ell}: \cP(m,k)\times \cP(n,r) \to \cP(m+n-\ell, k+r-\ell) 
\end{equation} 
for subsets $\{ i_1,\ldots, i_\ell \}\subset \{ 1, \ldots, k \}$ and $\{ j_1,\ldots, j_\ell \} \subset \{ 1, \ldots, n \}$ with
$i_s < i_{s+1}$ and $j_s< j_{s+1}$ for $s=1,\ldots,\ell-1$.  These composition operations satisfy
associativity axioms and unity axioms (see \cite{Kock}, \cite{Markl}, \cite{Valette}). 
The properad is symmetric if, moreover, 
there is an action of $\Sigma_n\times \Sigma_m$ on $\cP(n,m)$ with respect to which the composition
maps are bi-equivariant. 

\smallskip

We think of the objects in $\cP(n,m)$, in a properad $\cP$, as computational architectures with
$n$ inputs and $m$ outputs, with the composition operations $\circ^{i_1,\ldots, i_\ell}_{j_1,\ldots, j_\ell}$ that
graft the subset of outputs $\{ i_1,\ldots, i_\ell \}$ of the first machine to the subset of inputs $\{ j_1,\ldots, j_\ell \}$
of the second machine, in the given order, see Figure~\ref{ImageTryPropFig}. 
In the special case where we restrict to computational architectures
with a single output, we obtain the case of an operad.
One can also consider non-symmetric properads and operads (namely without the
condition of equivariance with respect to the symmetric group actions). Indeed, 
we will focus on examples that are non-symmetric. Thus, in the following, when we refer to
operads and properads, we will not assume the symmetric condition unless
explicitly stated. 

\begin{figure}
 \begin{center}
 \includegraphics[scale=0.45]{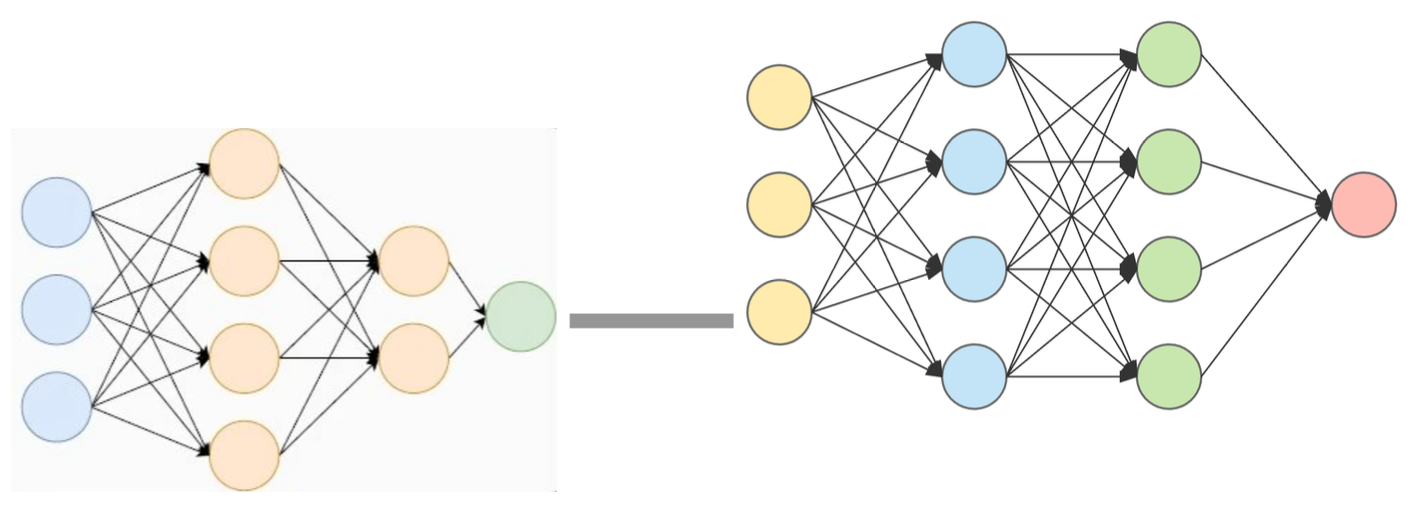}  
 \end{center} 
 \caption{Example of operad composition $\circ_3: \cO(3)\times \cO(3)\to \cO(5)$ connecting the single output of the first object (machine) to the third input of the second one. The inputs of the resulting object (machine) are the inputs of the first one together with the remaining inputs of the second.
\label{ImageTryOpFig}}
 \end{figure}

\smallskip

An operad in $\text{Cat}$  is a collection $\cO=\{ \cO(n) \}_{n\in \N}$ of small categories with composition functors
\begin{equation}\label{operad}
 \circ_j: \cO(m)\times \cO(n) \to \cO(m+n-1),  
\end{equation} 
subject to associativity, unity, and symmetry constraints as above. 

\smallskip

In the case if a symmetric operad one also requires
the existence of actions of $\Sigma_n$ on $\cO(n)$ with respect to which the composition maps are equivariant.

\smallskip

In the case of operads, 
the objects of $\cO(n)$ have $n$ inputs and one output and the composition operation $\circ_j$ connects the
output of the first machine to the $j$-th input of the second, thus giving a new machine with $n+m-1$ inputs and
a single output, see Figure~\ref{ImageTryOpFig}.

\smallskip

In particular we say that a unital symmetric monoidal category $\cC$ carries a properad 
structure when  
there is a family of full subcategories $\cC(n,m)$ for $n,m\in \N$
with the properties:
\begin{itemize}
\item ${\rm Obj}(\cC)=\sqcup_{n,m\in \N} {\rm Obj}(\cC(n,m))$;
\item the monoidal structure $(\otimes, {\mathbb I})$ satisfies
\begin{equation}\label{monoidProp}
 \otimes: \cC(m,k) \times \cC(n,r) \to \cC(m+n, k+r)\, ; 
\end{equation} 
\item the family $\{ \cC(n,m) \}_{n,m\in \N}$ is a properad in ${\rm Cat}$.
\end{itemize}

\smallskip

It is customary to also require additional compatibility conditions between the
monoidal structure and the properad structure. For instance, in the properad case 
we can consider compatibility conditions of the form
\begin{equation}\label{compatProp}
 (A \otimes B)\circ^{\cI\sqcup \cI'}_{\cJ\sqcup \cJ'} (C\otimes D) 
= (A\circ^\cI_\cJ C)\otimes (B\circ^{\cI'}_{\cJ'} D)\, , 
\end{equation}
for $A\in \cP(m,k)$, $C\in \cP(n,r)$, $B\in \cP(m',k')$ and $D\in \cP(n',r')$
and with $\cI=\{ i_1,\ldots, i_\ell \}\subset \{ 1,\ldots, k \}$, $\cJ=\{ j_1,\ldots, j_\ell \}\subset \{ 1,\ldots, n \}$,
$\cI'=\{ i'_1,\ldots, i'_{\ell'} \}\subset \{ 1,\ldots, k' \}$, $\cJ'=\{ j'_1,\ldots, j'_{\ell'} \}\subset \{ 1,\ldots, n' \}$.
Here the ``$=$" sign in \eqref{compatProp} should in general indicate isomorphism in the
category $\cP$ rather than strict equality.

\smallskip

We will discuss in \S \ref{IDAGsec} an example where compatibility
conditions of the form \eqref{compatProp} hold.

\smallskip

In the case of an operad structure we similarly have the following setting.

\smallskip

Let $\cC$ be a symmetric monoidal category with full subcategories $\cC(n)$ for $n\in \Z_{\geq 0}=\N\cup \{ 0 \}$ with
the following properties:
\begin{itemize}
\item ${\rm Obj}(\cC)=\sqcup_{n\in \Z_{\geq 0}} {\rm Obj}(\cC(n))$;
\item the $\{ \cC(n) \}_{n\in \N}$ form an operad in ${\rm Cat}$ with unital associative composition functors
$$  \circ_j: \cC(m)\times \cC(n) \to \cC(m+n-1)\, ; $$
\item $\cC(0)=\{ {\mathbb I} \}$, with ${\mathbb I}$ the unit of the monoidal structure $(\otimes, {\mathbb I})$, and 
for all $n,m\in \Z_{\geq 0}$ 
\begin{equation}\label{otimesOper}
 \otimes: \cC(n) \otimes \cC(m) \to \cC(n+m)\, . 
\end{equation} 
\end{itemize} 

 \smallskip
 
 First observe that if the operad structure on the $\cC(n)$ is induced from a properad 
 by taking the single-output $\cP(n,1)$, the monoidal structure \eqref{otimesOper} necessarily 
 differs from the monoidal structure \eqref{monoidProp}, which would instead give
 \begin{equation}\label{monoid1Prop}
  \otimes: \cP(n,1)\times \cP(m,1) \to \cP(n+m, 2)\, . 
 \end{equation} 
 Indeed, in specific cases (like the one we will be discussing in \S \ref{IDAG0sec} below) 
 one can obtain a monoidal structure satisfying \eqref{otimesOper} 
 from \eqref{monoidProp} using the monoidal structure of \eqref{monoid1Prop}
 followed by an ``identification of the two outputs" in \eqref{otimesOper}. We will
 discuss this more explicitly in \S \ref{IDAG0sec}.
 
 \smallskip
 
 Note that in this case the 	compatibility conditions between the monoidal
 structure and the prooperad composition of \eqref{compatProp} 
 also do not directly induce compatibility conditions of the same form on the
 operad. However we do obtain an induced compatibility condition
 of the form
  \begin{equation}\label{compatOper}
  A \circ_j (C\otimes D) =\left\{ \begin{array}{ll} (A\circ_j C)\otimes D & j=1,\ldots, m \\
  C \otimes (A\circ_{j'} D) & j'=j-m=1,\ldots, k, \, 
  \end{array}\right. 
  \end{equation}
 for all $A\in \cC(m)$, $C\in \cC(n)$, $D\in \cC(k)$ with $m,n,k\in \N$. Again the ``$=$" sign in general
 denotes isomorphism in the category $\cC$ rather than strict equality.

\smallskip

This will be useful in
the specific example we discuss in \S \ref{IDAG0sec} below. 

\smallskip

Note that we have the unit of the operad composition, ${\bf 1}\in \cC(1)$, satisfying
${\bf 1}\circ_j A = A= A\circ_j {\bf 1}$, for all $A\in \cC(n)$ and $n\in \N$, and
the unit of the monoidal structure ${\mathbb I}\in \cC(0)$ satisfying
${\mathbb I}\otimes A = A= A \otimes {\mathbb I}$, for all $A\in \cC(n)$ and $n\in\Z_{\geq 0}$.
In general we do not assume a relation between these two objects, but we will discuss
explicitly how they are related in the example we discuss in \S \ref{IDAG0sec} below. 

\medskip
\subsubsection{Compositional network summing functors}

Let $\cC$ be a category of resources given by a unital symmetric monoidal category with a properad structure $\{ \cC(n,m) \}$ as
above. Let $G$ be a finite directed graph. A compositional $\cC$-valued network summing functor
on $G$ consists of an assignment of objects of $\cC$ to subnetworks of $G$, functorial under inclusions,
with the property that the objects $\{ \Phi(v) \}_{v\in V(G)}$ of $\cC$ associated to the vertices of $G$
belong to ${\rm Obj}(\cC(\deg^{in}(v), \deg^{out}(v)))$ with $\deg^{in/out}(v)$ the in and out degrees
of the vertex $v$, and that for any $G'\subseteq G$ the object $\Phi(G')$ is obtained from the objects
$\Phi(v)$ for $v\in V(G')$ by applying the composition operations $\circ^{i_1,\ldots, i_\ell}_{j_1,\ldots, j_\ell}$ 
of the properad, as in \eqref{properad}, according to the gluing of input and output half-edges of the corollas 
of the vertices $v\in V(G')$ to form the edges of $E(G')$. We will return to this definition in more detail
in \S \ref{LeakSec} below. We will also describe more explicitly notation and assumptions about the
networks $G$ in \S \ref{GraphSec}. 

\smallskip

We write $\Sigma_\cC(G)$ for the category of compositional $\cC$-valued network summing functor
on $G$, defined as above, with invertible natural transformations. By the compositional property, the summing functors
are completely determined by the objects $\{ \Phi(v) \}_{v\in V(G)}$ and the invertible natural transformations by
isomorphisms of these objects.

\medskip
\subsection{Hopfield equations and threshold linear networks}\label{HopSec}

The classical Hopfield equations were introduced to model memory storage and retrieval in
neuronal networks, adapting ideas from the statistical physics of spin glasses, \cite{Hopfield}.

\smallskip

In this original form, the variables in the equation are binary $x_j=\pm 1$, describing the on/off firing
status of individual neurons over a time interval $\Delta t$ of fixed size. The equations take the form
$$ x_i(n+1)={\rm sign} \left( \sum_j W_{ij} x_j(n) \right)\, , $$
where $W_{ij}$ is a weight matrix that describes the strength of the connectivity between 
different neurons. A model of memory storage and retrieval is obtained by choosing a weight
matrix determined by a given set of patterns (binary strings $\pi^a_i$, $a=1,\ldots, M$)
one wants to store in the network (Hebbian learning rule)
$$ W_{ij}=\frac{1}{N} \sum_{a=1}^M \pi^a_i \pi^a_j \, . $$
The convergence of the dynamics to the minima of the energy functional
$$ \cE=-\sum_{ij} W_{ij} x_i x_j $$ corresponds to convergence to the stored
patterns whenever the input data are sufficiently close to one of the patterns,
so that the stored memory can be reconstructed from partial data, see
\cite{Amit}, \cite{Gerstner} for a more detailed account.

\smallskip

A continuous variables version of the Hopfield equations (see \cite{Hopfield2})
was developed to model the firing rates of neurons, which are now described by
real variables $x_j$ in the equations, which take the form
\begin{equation}\label{ContHopfEq2}
\frac{dx_i}{dt} = - \frac{x_i}{\tau} +  \sigma\left( \sum_j W_{ij} x_j + \theta_i \right) \, ,
\end{equation}
with a non-linear function $\sigma$ used to ensure that the input firing rates are 
non-negative. A ``bias term" $\theta_i$ is also introduced in this form of the equation
to model the presence of an external input (an injected current). The first term
on the right-hand-side of the equation \eqref{ContHopfEq2} is referred to as
``leak term" and is responsible for describing a synaptic current that decays 
at an exponential rate $\exp(-t/\tau)$. The weight matrix $W_{ij}$ describes the
excitatory or inhibitory action (depending on the sign of the matrix entries) 
of neurons on other neurons, while the leak term describes the inhibitory action
of a neuron on itself. 

\smallskip

In particular, the non-linearity $\sigma$ can be taken of the threshold-linear 
form $$\sigma(x)=(x)_+:=\max\{ x, 0 \}.$$ In this case the Hopfield equations
\eqref{ContHopfEq2} are also known as ``threshold linear networks". 

\medskip
\subsubsection{Inhibitory-excitatory balance}\label{InexBalSec} 

The matrix entries $W_{ij}$ of the weight matrix in \eqref{ContHopfEq2} are real numbers
can have either positive or negative sign. This is the reason why one thinks of these
equations as the analog of a statistical physics spin glass model rather than an Ising
model where the interaction weights are positive and energetically favor spin alignment.
The negative and positive signs of the entries $W_{ij}$
here model inhibitory and excitatory interactions between
neurons. 

\smallskip

Experimental evidence in neuroscience shows that inhibitory and excitatory interactions
in the cerebral cortex are well balanced during resting states and during sensory processing,
with a balanced condition that occurs either in the form of an overall global balance 
or a temporal balance. It is shown in \cite{Brunel} that perturbing the global 
inhibitory-excitatory balance condition in favor of either excitation dominating
inhibition or viceversa has significant effects on the behavior of the resulting dynamics, 
while in \cite{Vreeswijk} it is shown that inhibitory-excitatory balance can emerge
from the dynamics (in a mean field theory model) as a stationary state of large networks.
Moroever, it is shown in \cite{ZhouYu} that the inhibitory-excitatory balance condition
may be a fundamental mechanism underlying efficient neural coding.

\smallskip

In this paper we show that a global inhibitory-excitatory balance condition arises
naturally from the categorical form of the Hopfield equations introduced in
\cite{ManMar}, imposed by the categorical form of the threshold nonlinearity
of the equations. We also show the effects on the dynamics of the violation
of the inhibitory-excitatory balance.

\smallskip
\subsection{Dynamics in categories of summing functors}\label{DynSummingSec}

In \cite{ManMar} categorical forms of the Hopfield network equations were introduced, motivated by 
a finite difference version of the classical Hopfield equations. The latter 
can be written either in the form
\begin{equation}\label{discrHopf1}
 \frac{x_i(t+\Delta t)-x_i(t)}{\Delta t} = (\sum_j W_{ij} x_j(t) +\theta_i)_+\, ,  
\end{equation} 
or with an additional ``leak term" $-x_i(t)$ as
\begin{equation}\label{discrHopf2}
 \frac{x_i(t+\Delta t)-x_i(t)}{\Delta t} =-x_i(t) + (\sum_j W_{ij} x_j(t) +\theta_i)_+\, , 
\end{equation} 
where the non-linear threshold on the right-hand-side is given by $(\cdot)_+=\max\{ 0, \cdot \}$,
and the weight matrix $W_{ij}$ and bias term $\theta_i$ are as discussed in \S \ref{HopSec} above.
Note that \eqref{discrHopf2} is the discretized finite-difference version of the differential equation
\eqref{ContHopfEq2} (taken with $\tau=1$). 

\smallskip

An analog of these equations is obtained by replacing the real-valued variables $x_j(t)$ describing neuron
firing rates with variables in the category of network summing functors describing assignments of resources
of type $\cC$ to the network, with the matrix $W$ of weights replaced by a matrix $T$ of
endofunctors of the category of
resources. The resulting categorical Hopfield network equations proposed in \cite{ManMar} take the form
\begin{equation}\label{catHopf1}
 X_v(n+1)= X_v(n)\oplus \left( \oplus_{e\in E\, : \, s(e)=v} T_{e}(X_{t(e)}(n)) \oplus \Theta_v \right)_+ 
\end{equation} 
or possibly 
\begin{equation}\label{catHopf2}
 X_v(n+1)= \left( \oplus_{e\in E\, : \, s(e)=v} T_{e}(X_{t(e)}(n)) \oplus \Theta_v \right)_+ \, ,
\end{equation} 
We refer to the first case \eqref{catHopf1} as ``with leak term" and the second case
\eqref{catHopf2} as ``without leak term".  

\smallskip

Here the $T_e$ for $e\in E(G)$ (or $T_{vv'}$ for $v,v'\in V(G)$)
are a collection of endofunctors of $\cC$ and $\Theta$ is an assigned summing functor
(specified by the objects $\Theta_v$ in $\cC$), while the variables $X_v(n)$ of the equation
are a sequence of objects $X(n)\in \Sigma_\cC(G)$, each determined by the objects $\{ X_v(n) \}_{v\in V}$ of $\cC$. The nonlinear threshold is now defined in terms of convertibility conditions in the category $\cC$
of resources, where for $C\in {\rm Obj}(\cC)$ we set $( C )_+= C$ if $[C]\succeq 0$ and $(C)_+=0$ 
(the unit of the monoidal structure of $\cC$) otherwise. 

\smallskip

The endofunctors $T$ of the category of resources are assumed to
preserve the unital symmetric monoidal structure of $\cC$, 
although the non-linear threshold does not. 

\smallskip

We can also write the equations \eqref{catHopf1} \eqref{catHopf2} above in the form
\begin{equation}\label{catHopfVert}
\begin{array}{rl}
 X_v(n+1)= & X_v(n)\oplus \left( \oplus_{v'\in V} T_{vv'}(X_{v'}(n)) \oplus \Theta_v \right)_+ \ \ \text{ or } \\
 X_v(n+1)= & \left( \oplus_{v'\in V} T_{vv'}(X_{v'}(n)) \oplus \Theta_v \right)_+ \, ,
 \end{array}
\end{equation} 
where we identify 
$$ T_{vv'}=\oplus_{e\in E\, : \, s(e)=v,\, t(e)=v'} T_e\, , $$
with the monoidal sum applied pointwise $(\oplus_e T_e)(X):=\oplus_e T_e(X)$.
In the case where the graph does not have multiple edges, $T_{vv'}(X_{v'})=T_e(X_{v'})$ if there
is an edge between $v$ and $v'$ and is $0$ otherwise. Thus, we 
can use either the form \eqref{catHopf1} and \eqref{catHopf2} or the form \eqref{catHopfVert}
of the equations.  

\smallskip

It it important to note that in equations such as \eqref{catHopf1}, \eqref{catHopf2}, \eqref{catHopfVert},
since we are dealing with objects in a category, the ``$=$" sign of the equation can be interpreted as
the {\em existence of an isomorphism} in the given category between the objects in the left-hand-side
and the right-hand-side of the equation. In the case of a small category, one can also interpret the
``$=$" sign in the equation in the stricter sense of {\em equality} rather than isomorphism of objects,
but equality in general is not as good a notion in a category theory setting as isomorphism, so it is 
preferable to interpret the equation as requiring isomorphism. As we discuss briefly in
\S \ref{FixedPtSec}, these two different intepretations of the equations lead to different notions of
fixed points and periodic orbits of the dynamics. 

\smallskip

In this paper we will discuss  
these two cases (with and without leak term) 
and their different behavior in a toy model example with a particular
choice of the category $\cC$ of resources. 
We write the equations as above, 
in terms of variables $X_v$ at vertices of $G$ rather than edges as in \cite{ManMar}
because of our specific choice of category of network summing functors, 
given by the compositional network summing functors defined above,
which are determined by objects of $\cC$ assigned to vertices rather than edges.

\smallskip

The goal of the present paper is to illustrate more concretely the properties of these
categorical Hopfield equations in a specific example motivated by a
computational model of neurons developed in \cite{BeSeLo},
with our category $\cC$ of (computational) resources given by a category of 
deep neural networks (DNNs).

\medskip
\subsubsection{Fixed points in categories}\label{FixedPtSec}

Before discussing specific examples of the categorical dynamics introduced
above, we recall some general facts about fixed points of endofunctors that
will be useful in the following. 

\smallskip

An important first step in the study of dynamical systems is the structure of fixed points.
In a categorical setting like the one we are considering here,
this involves discussing fixed points of endofunctors in categories.
The same setting can be adapted to study the existence of periodic
orbits, as fixed points of powers of the same endofunctor.

\smallskip

Given a category $\cC$ and an endofunctor $F: \cC \to \cC$, a fixed point for
$F$ is a pair $(X,\alpha)$ of an object $X\in {\rm Obj}(\cC)$ and an isomorphism
$\alpha: X \to F(X)$ in ${\rm Mor}_{\cC}(X,F(X))$.

\smallskip

Fixed points of an endofunctor $F$ of a category $\cC$ form a category ${\rm Fix}(F)$ with
objects the fixed points $(X,\alpha)$ and morphisms in ${\rm Mor}_{{\rm Fix}(F)}((X,\alpha),(Y,\beta))$
given by commutative diagrams
$$ \xymatrix{ X \ar[r]^\alpha \ar[d]_f & F(X) \ar[d]^{F(f)} \\ Y \ar[r]_\beta & F(Y) } $$
with a morphism $f\in {\rm Mor}_\cC(X,Y)$. There is a forgetful functor
${\rm Fix}(F) \to \cC$ that maps $(X,\alpha)\mapsto X$ and a diagram as above
to the morphism $f: X\to Y$. 

\smallskip

Note that the category ${\rm Fix}(F)$ can be empty
if the endofunctor $F$ has no fixed points. An example of an
endofunctor with no fixed points is the functor $\cP: {\rm Sets} \to {\rm Sets}$
with $\cP(X)=2^X$ the power-set functor, since there cannot be a
bijection between a set and its set of parts. 

\smallskip

A fixed point $(X,\alpha)$ of an endofunctor $F$ on a small category $\cC$ i
s {\em strict} if $F(X)=X$ with $\alpha$ the identity morphism. The existence of strict fixed
points is in general a much more restrictive condition. However, the following
criterion (Proposition~2.6 of \cite{Luzh}) is especially useful:
An endofunctor $F: \cC\to \cC$ has a strict fixed point $F(X)=X$ if
the induced continuous map $|F|: |\cN\cC|\to |\cN\cC|$ on the topological
realization of the nerve $\cN\cC$ of the category $\cC$ has a fixed point.

\smallskip

A point of order $n$ of an endofunctor $F$ is a fixed point of the endofunctor $F^n$
that is not a fixed point of any other $F^k$ with $k<n$,
namely a pair $(X,\gamma)$ of an object $X$ and an isomorphism $$\gamma: X \stackrel{\simeq}{\to} F^n(X), $$, 
such that $X\not\simeq F^k(X)$ for any $k<n$. A periodic orbit of order $n$ is a set $\{ (X_1, \gamma_1), \ldots, (X_n,\gamma_n) \}$ of points of
order $n$ where 
$$ \gamma_{i+1}: X_{i+1} \stackrel{\simeq}{\to} F(X_i)  \ \ \text{ and } \ \  \gamma_1: X_1 \stackrel{\simeq}{\to} F(X_n) . $$

\smallskip

In a periodic orbit of order $n$, in particular,  $(X_1,  \alpha_1=F^{n-1}(\gamma_n)  \cdots F(\gamma_2) \circ \gamma_1)$ is a point of order $n$ 
and so are all the $(X_i, \alpha_i)$ with the isomorphism $\alpha_i: X_i \to F^n(X_i)$ obtained analogously. Points of order $n$ of an endofunctor $F$ 
form a subcategory of the category ${\rm Fix}(F^n) \subset \cC$, which we denote by ${\rm Fix}_n(F)$. Note that it is a subcategory because 
${\rm Fix}(F^n)$ also contains points fixed by $F^n$ that are of some order $k < n$. 
Orbits of order $n$ also form a category with objects $\{ (X_1, \gamma_1), \ldots, (X_n,\gamma_n) \}$ as above and with morphisms
$(f_1,\ldots, f_n)$ with $f_i\in {\rm Mor}_{\cC}(X_i, X_i')$ with commutative diagrams
$$ \xymatrix{ X_{i+1} \ar[r]^{\gamma_{i+1}} \ar[d]_{f_{i+1}} & F(X_i) \ar[d]^{F(f_i)} \\ X'_{i+1} \ar[r]_{\gamma'_{i+1}} & F(X'_i) }\, . $$
We denote the category of orbits of order $n$ by $\cO_n(F)$, or by $\cO_{n,\cC}(F)$ if we want to make the ambient category
explicit.

\section{Categories of DNNs and Hopfield dynamics}

In the rest of this paper we will focus on a case where the category of resources $\cC$
is a category of deep neural networks. The variables in the categorical Hopfield equations
will therefore assign to nodes of a given networks some DNN architectures given by
objects in this category $\cC$ of DNNs. This means that we want to model the
individual neurons of a neuronal system with DNNs. This choice is motivated by
some recent results in computational neuroscience, such as \cite{BeSeLo}, where
it is shown that individual cortical neurons can indeed be modeled by certain DNNs.

\smallskip

In \cite{BeSeLo}, a computational model of cortical neurons was obtained in terms of deep neural networks (DNNs)
trained to repilcate the input/output function of biophysical models of cortical neurons at millisecond (spiking) resolution. 
Thus, for instance, a a layer 5 cortical pyramidal cell (L5PC) is modeled by a temporally convolutional DNN with five to 
eight layers, while for simpler models of cortical neurons (not including NMDA receptors) a fully connected neural network 
with one hidden layer (consisting of 128 hidden units) suffices. In particular, it is shown in \cite{BeSeLo} that the weights
of the resulting DNN model synapses, in the sense that the learning process automatically produces weight matrices 
of the model DNNs  that incorporate both positive and negative weights corresponding, respectively, to excitatory and 
inhibitory inputs in the integrate-and-fire model of the neuron.

\smallskip

Motivated by this computational model of neurons, we specialize our discussion of categorical Hopfield dynamics
to a particular example, where our choice of a category $\cC$ of resources is given by a category of DNNs.

\subsection{Monoidal categories of DNNs}

We want to organize DNNs in a categorical structure that has both a symmetric monoidal
structure (which in our setting corresponds to decompositions into independent subsystems) and
a structure of compositionality, which describes how outputs of some processes can be fed to inputs of others. 
The setting we describe here is closely related to categorical models of DNNs already
developed in \cite{FioCa}, \cite{FoSpiTu}, \cite{Ganchev}.

\medskip
\subsubsection{Notation and assumptions about graphs}\label{GraphSec}

We discuss briefly some general notation, terminology and assumptions
about graphs that we will be using in the rest of the paper. In particular,
here we briefly recall and compare different categorical notions of directed
graphs and morphisms of directed graphs and the respective relevance for
the setting we will consider in the following. 

\smallskip

A first categorical formulation of directed graphs is based on the following idea.
Directed finite graphs are functors from a category ${\bf 2}$ that has two objects $\{ E, V \}$
and two non-identity morphisms $s,t: E \to V$ (the source and target morphisms) to
the category of finite sets $\cF$. 
Morphisms of directed finite graphs are natural transformations of these functors.
Thus, the category of directed finite graphs is the category
of functors
$$ \cG ={\rm Func}({\bf 2}, \cF)\, . $$
This means that a directed finite graph $G$ assigns to the two objects $E,V$ two
sets, which we denote by $E(G)$ and $V(G)$, respectively. These are 
the set of directed edges and the set of vertices of the graph $G$. The functor also 
to the two morphisms $s,t$ corresponding maps of sets, which we still write with the same
notation, $s,t: E(G)\to V(G)$. These 
map an edge $e\in E(G)$ to its source and target vertices $s(e), t(e)\in V(G)$. A
natural transformation $\varphi: G \to G'$ then consists of two maps of sets
$\varphi_E: E(G) \to E(G')$ and $\varphi_V: V(G) \to V(G')$ that are compatible with
the source and target morphisms, namely such that the following diagrams commute:
\begin{equation}\label{stnattrans}
 \xymatrix{ E(G) \ar[r]^{\varphi_E} \ar[d]_s & E(G') \ar[d]^s \\
V(G) \ar[r]_{\varphi_V} & V(G') } \ \ \ \text{ and } \ \ \ 
\xymatrix{ E(G) \ar[r]^{\varphi_E} \ar[d]_t & E(G') \ar[d]^t \\
V(G) \ar[r]_{\varphi_V} & V(G')\, . }
\end{equation}

\smallskip

Note, however, that this version of morphisms of directed graphs does not
include contraction of edges (which would correspond to mapping an edge to a vertex).
Since contractions will play a useful role in our setting, one needs to
slightly modify the category of directed graphs described above to allow for these
additional morphisms. This can be done by considering a source category $\tilde{\bf 2}$
with the same objects $\{ E, V \}$ and the source and target morphisms, but with an 
an additional morphism $\epsilon: V \to E$, with relations 
$s\circ \epsilon=t \circ \epsilon ={\rm id}_V$, which represents a looping edge
at a vertex, see  \cite{Brown}. Then a functor $G$ also determines a map of sets
$\epsilon: V(G) \to E(G)$. 
A morphism $\varphi: G \to G'$ will then also satisfy the commutativity of 
$$  \xymatrix{ E(G) \ar[r]^{\varphi_E} & E(G') \\
V(G) \ar[u]^\epsilon \ar[r]_{\varphi_V} & V(G')\ar[u]_\epsilon  } \, . $$
Given an element $\epsilon(v')\in E(G')$, for some $v'\in V(G')$, the
set of edges $e\in E(G)$ that are in the preimage $\varphi_E^{-1}(\epsilon(v'))$
describe the edges of $G$ that are contracted to the vertex $v'$ in $G'$.
We denote the resulting category of directed graph by
$$ \tilde\cG={\rm Func}(\tilde{\bf 2}, \cF)\, . $$

\smallskip

In the following, we will consider subcategories of the category $\tilde\cG$ of directed finite
graphs. For instance, in \S \ref{IDAGsec} we will consider directed graphs that
are acyclic (no directed loops) and with the set of vertices $V(G)=I\cup H \cup O$
consisting of inputs $I$, hidden nodes $H$, and
outputs $O$, with morphisms induced from the morphism in $\tilde\cG$.

\smallskip

In order to better describe the operations of plugging outputs of one
directed graph into inputs of another, it is convenient to also use another
description of directed graphs in terms of flags (half-edges) rather than
edges, so that the inputs and outputs can be marked by inward/outward
oriented flags (also called ``external edges" in the physics terminology).
In this setting one things of edges of a graph (``internal edges" in
the physics terminology) as matched pairs of half-edges while the
external edges are inputs and outputs that can be matched with
the inputs/outputs of another graph. In categorical terms, we can
use the following formulation. 

\smallskip

Consider a category $\tilde{\bf 2}^{i/o}$ that has a set of objects of the form
$\{ V, E, F_i ,F_o \}$ and that has non-identity morphisms
\begin{equation}\label{VEIO}
 E \stackrel{f_i}{\rightarrow} F_i \stackrel{t}{\rightarrow} V \stackrel{s}{\leftarrow} F_o \stackrel{f_o}{\leftarrow} E \, .
\end{equation} 
as well as the $\epsilon$ morphism described above with 
$t\circ f_i\circ \epsilon=s\circ f_o\circ \epsilon={\rm id}_V$.
A finite directed graph with external edges is a functor $G: \tilde{\bf 2}^{i/o}\to \cF$ 
with the property that the morphisms $f_i,f_o$ are mapped to injective maps.
Thus, graphs described in terms of flags are a subcategory of the
category of functors in ${\rm Funct}(\tilde{\bf 2}^{i/o}, \cF)$
and morphisms of directed graphs are natural transformations of these functors. 

\smallskip

Here we have again a set of vertices $V(G)$ and a set of (internal) edges $E(G)$,
with source and target maps $s\circ f_o,t\circ f_i: E(G) \to V(G)$. Moreover, we now have also 
sets $F_i(G)$ and $F_o(G)$ of incoming/outgoing flags (half-edges oriented
to/from the vertex), with $t: F_i(G)\to V(G)$ and $s:  F_o(G)\to V(G)$ the
respective boundary morphisms that assign to a flag its boundary vertex. 
The morphisms $f_i: E(G)\to F_i(G)$ and $f_o: E(G)\to F_o(G)$ assign to an
edge of $G$ the two flags (half-edges), respectively attached to source and target vertex,
that are matched together to form the edge. The set of external edges (input/output flags) 
of $G$ is then
$$ E_{ext}(G)=(F_i(G)\smallsetminus f_i(E(G))) \sqcup (F_o(G)\smallsetminus f_o(E(G))) \, . $$

\smallskip

In the following, we will consider finite directed graphs where a given vertex has either
no incident external edges or at most one (either incoming or outgoing). Thus, specifying
the subsets $I,O$ (input and output vertices) of $G$, which are exactly the vertices that
have an external edge, is equivalent to completely describing the directed graph in terms of flags. 
We will switch between the notation with input and output vertices $I,O$ and the notation
with flags wherever it is more convenient. 

\smallskip

In the following we will consider directed graphs in two different roles. An
underlying graph, for which we will use the notation $\fG$, which is the
network over which the categorical Hopfield equations take place. We will
not make any specific assumptions about this graph other than requiring that it is 
finite and directed. We will also consider directed graphs $G_v$ that (together
with weights assigned to their internal edges) are
objects in a category of resources, assigned to the vertices $v\in V(\fG)$
by a summing functor. There will be additional assumptions on these
graphs $G_v$: that they are acyclic and that each input vertex has a single 
incoming external edge and each output vertex has a single outgoing external edge 
(IDAG category of \S \ref{IDAGsec}).
We will also consider a class of such graphs with the additional requirements
that there is only one output vertex and that 
all vertices have a single outgoing (internal or external) edge ($\text{IDAG}^0$ category
of \S \ref{IDAG0sec}).

\medskip
\subsubsection{Category of IDAGs}\label{IDAGsec}

We consider, as in \cite{FioCa}, ``interfaces directed acyclic graphs" (IDAGs). These are finite acyclic graphs $G=(V,E)$ 
with vertex set $V$ subdivided into three subsets $V=I\cup H \cup O$, respectively indicating
inputs, hidden nodes, and outputs. 
When $\# V\geq 2$ we assume these three subsets of vertices are disjoint, $V=I\sqcup H \sqcup O$.
However, in the case of a single vertex $V=\{ v \}$, we also allow for the possibility that $v$ may be
both an input and an output, namely $I=O=V$ and $H=\emptyset$. 
Vertices in $I$ have only outgoing edges and vertices in $O$ have
only incoming edges, namely source and target maps satisfy $s: E\to I\sqcup H$ and $t: E \to H\sqcup O$. 
The special case with of a single vertex that is both input and output will be needed
to describe the unit of the compositional (properad) structure. 

\smallskip

One can also formulate the data of IDAGs in terms of flags (half-edges), so that each input vertex
has a single incoming half-edge (external input) and outgoing edges (to hidden nodes and outputs), 
each output vertex has a single outgoing half-edge
(external output) and incoming edges (from hidden nodes or inputs).
Since the external edges are completely determined by the assigment of input and output vertices,
we will describe the graphs in terms of the data of vertices $V=I\cup H \cup O$ and 
internal edges $E$, and we will only mention flags when needed. 

\smallskip

We view the category $\text{IDAG}$ with these objects as a subcategory of the
category of directed graphs $\tilde\cG$ described in \S \ref{GraphSec}, 
with the induced morphisms. 

\smallskip

The objects of the category of IDAGs we consider here are the same as the objects in 
the category of arrows of the category considered in \cite{FioCa}. Indeed in 
\cite{FioCa} one considers a category with objects
given by finite sets and morphisms ${\rm Mor}(I,O)$ given by the IDAGs
with input set $I$ and output set $O$, with composition given by concatenation. 
At the level of morphisms these categories do not agree, however, as in the
category of arrows a morphism between two IDAGs $G,G'$ is given by a pair 
of IDAGs $H,H'$ with compositions $H\circ G=G'\circ H'$, while morphisms in
our category of IDAGs consist of homomorphisms 
$\varphi: G \to G'$ of directed graphs (which map inputs to inputs and 
outputs to outputs). 

\smallskip

The compositionality structure we consider 
on $\text{IDAG}$ is given by a {\em properad structure} defined as in \S \ref{OpPropSec} with 
$\cP(n,m)=\text{IDAG}(n,m)$ given by the full subcategory of $\text{IDAG}$ with objects the IDAGs with
$n$ input vertices and $m$ output vertices and with composition rules $\circ^{i_1,\ldots, i_\ell}_{j_1,\ldots, j_\ell}$
given by identifying the set of output vertices $\{ i_1,\ldots, i_\ell \}$ with the set of input vertices 
$\{ j_1,\ldots, j_\ell \}$, thus turning them into hidden nodes of the resulting IDAG, which is therefore in
$\text{IDAG}(m+n-\ell, k+r-\ell)$.

\smallskip

It is customary to consider on categories of directed graphs a monoidal structure $\oplus : \text{IDAG} \times \text{IDAG} \to \text{IDAG}$ given by
the disjoint union of directed graphs (which correspond to independent subsystems), with unit $0$ given by the empty graph. 

\smallskip

This monoidal structure satisfies the compatibility condition \eqref{compatProp} with the properad
composition. Indeed, if $\cI$, $\cJ$ and $\cI'$, $\cJ'$ are sets of indices as in \eqref{compatProp} 
then for any objects $G_i$, $i=1,\ldots,4$ with $G_1\in \text{IDAG}(m,k)$, $G_2\in \text{IDAG}(m',k')$, $G_3\in \text{IDAG}(n,r)$, $G_4\in \text{IDAG}(n',r')$, with $G_1\circ^{\cI}_{\cJ} G_3$ the object of
$\text{IDAG}(m+n-\ell, k+r-\ell)$ obtained by grafting the output subset $\cI$ of $G_1$ into the input subset $\cJ$ of $G_3$ and $G_2\circ^{\cI}_{\cJ} G_4$ the object of
$\text{IDAG}(m'+n'-\ell', k'+r'-\ell')$ obtained by grafting the 
output subset $\cI'$ of $G_2$ into the input subset $\cJ'$ of $G_4$ we have
$$ (G_1\circ^{\cI}_{\cJ} G_3) \otimes (G_2\circ^{\cI}_{\cJ} G_4) $$
given by the disjoint union of these two resulting graphs, which is the same as taking the properad
composition over the sets $\cI\sqcup \cI'$, $\cJ\sqcup \cJ'$ in the disjoint union of $G_1$ and $G_2$
and the disjoint union of $G_3$ and $G_4$, respectively, namely
$$ (G_1\otimes G_2)\circ^{\cI\sqcup \cI'}_{\cJ\sqcup \cJ'} (G_3\otimes G_4)\, , $$
hence the compatibility \eqref{compatProp}  is satisfied.

\smallskip

The unit ${\bf 1}\in \text{IDAG}(1,1)$ of the properad composition is given by
$${\bf 1}=( I=\{ o \}, H=\emptyset, O=\{ o \}, E=\emptyset )\, . $$ 

\smallskip
\subsubsection{Single output IDAGs}\label{IDAG0sec}

Since we are ultimately interested in DNNs, we can start by restricting the more general category of IDAGs described
above to the case of IDAGs with a single output node. We write $\text{IDAG}^o$ for the subcategory of $\text{IDAG}$
with objects consisting of IDAGs $G=(I,H,O,E)$ 
with set of vertices $V=I\cup H \cup O$ and set of edges $E$, 
and with the set $O$ of output nodes consisting of a single vertex $O=\{ o \}$.
We also assume that every vertex that is 
not an output has a single outgoing edge (to hidden nodes or to outputs). 
Morphisms are morphisms of IDAGs, which map output vertex to output vertex.

\smallskip

The restriction of the properad $\cP(n,m)=\text{IDAG}(n,m)$ to this subcategory gives an operad 
as in \S \ref{OpPropSec},  with $\cO(n)=\text{IDAG}^o(n)=\text{IDAG}(n,1)$.

\smallskip

On the category $\text{IDAG}^o$ we can also consider a monoidal structure defined in the following way. For $G=(I,H,\{ o \}, E)$ 
and $G'=(I',H',\{ o' \}, E')$ in  $\text{IDAG}^o$ we set
$$ G\oplus G'=(I\times \{ o' \} \sqcup \{ o \} \times I', H\times \{ o' \} \sqcup \{ o \} \times H', \{ (o,o')\}, E\times \{ o' \}\sqcup \{ o \} \times E')\, . $$

That is, the monoidal operation $\oplus$ is given by first taking a disjoint union and then identifying
the output vertices. 
The unit for the monoidal structure is the IDAG $0=(\emptyset, \emptyset, \{ o \}, \emptyset)$ consisting of a single output vertex $\{ o \}$
with $I=\emptyset$, $H=\emptyset$, $E=\emptyset$.  

\smallskip

This choice of monoidal structure is similar to the monoidal structure on the category of concurrent/distributed computing
architectures given by transition systems automata, introduced in \cite{WiNi95}   and considered in \cite{ManMar}.

\smallskip

As we discussed in \S \ref{OpPropSec} the compatibility condition \eqref{compatProp} between
the properad composition and the monoidal structure given by disjoint union induces a different
compatibility condition between the operad composition and the monoidal structure obtained by
taking disjoint union followed by identification of the outputs, as in \eqref{compatOper}. Indeed
it can be easily seen that \eqref{compatOper} is verified, by arguing as in the case of 
\eqref{compatProp} discussed above.

\begin{figure}
 \begin{center}
 \includegraphics[scale=0.35]{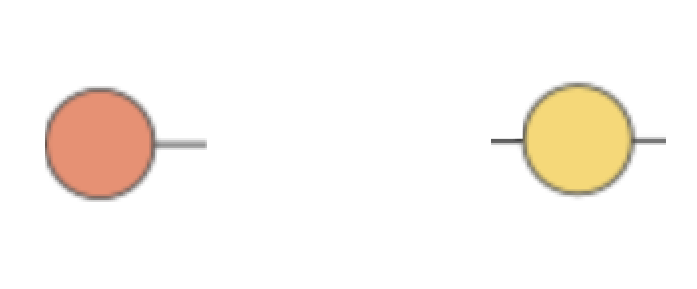}  
 \end{center} 
 \caption{The unit ${\mathbb I}=0 \in \text{IDAG}^0(0)$ of the monoidal structure 
 as a vertex with a single outgoing flag and the unit ${\bf 1}\in \text{IDAG}^0(1)$ of the operad
 structure as a vertex with a single input flag and a single output flag.
\label{UnitsFig}}
 \end{figure}

\smallskip

 It is in fact convenient to describe the unit for the monoidal structure ${\mathbb I}=0\in \text{IDAG}^0(0)$,  
given by the single output $0=(\emptyset, \emptyset, \{ o \}, \emptyset)$, in terms of flags, by viewing
the output vertex $\{ o \}$ as consisting of a pair $o=(v_o,f_o)$ 
of a vertex $v_o$ with an attached outward-oriented flag (half-edge) 
$f_o$, so as to be reminded of
the fact that this vertex has a non-empty output set. Note that we have $E=\emptyset$, as there is
no pair of flags forming an edge, see Figure~\ref{UnitsFig}.

\smallskip

The unit of the operad composition ${\bf 1}\in \text{IDAG}^0(1)$, on the other hand
consists of a single vertex $v$ together with two half edges $f_v, f'_v$, one incoming
and one outgoing at $v$, $t(f_v)=v$ and $s(f'_v)=v$, see Figure~\ref{UnitsFig}. The operadic composition  
$\circ: \text{IDAG}^0(k)\times \text{IDAG}^0(1)\to \text{IDAG}^0(k)$ plugs the
outgoing flag of the single output of an object $A$ of $\text{IDAG}^0(k)$ into the single input
flag $f_v$ of ${\bf 1}$, by gluing together the outgoing flag and the ingoing flag $f_v$
to an edge $e$, followed by the operation that contracts this edge, so that the vertex $v=t(e)$
is brought to coincide with the vertex $s(e)$ and only the outgoing flas $f'_v$ remains, while
the resulting set of edges (internal edges) of $A\circ {\bf 1}$ agrees with those of $A$
and we have an identification of these two objects. 
The composition $\circ_j: \text{IDAG}^0(1)\times \text{IDAG}^0(k)\to \text{IDAG}^0(k)$
is done similarly with the outgoing flag $f_v'$ matched with the $j$th incoming flag
of $A$ and the resulting edge $e$ then contracted to obtain an identification of
${\bf 1}\circ_j A$ and $A$.

\smallskip
\subsubsection{Weighted single output IDAGs}

We now modify the construction above of the category $\text{IDAG}^o$ in order to incorporate 
data of weights on the edges of the networks. The category $\text{WIDAG}^o$ of weighted single-output IDAGs
has objects given by pairs $(G,W)$ with $G=(I,H,\{ o \}, E)$ an IDAG with a single output and 
$W$ a weight matrix, namely a function $W: V\times V \to \R$ where $V=I\sqcup H\sqcup \{ o \}$ the
set of vertices, with the property that $W(u,v)=0$ whenever $\{ e\in E\,|\, s(e)=u,,\, t(e)=v \}=\emptyset$,
with $s: E\to I\sqcup H$ and $t: E \to H\sqcup O$ the source and target map.
Note that because the graphs in $\text{IDAG}$ are assumed to be acyclic, they in particular do not contain
looping edges, hence $W(v,v)=0$ for all $v\in V$. 
Morphisms $\varphi\in {\rm Mor}_{\text{WIDAG}^o}((G,W),(G',W'))$ are morphisms $\varphi: G \to G'$ 
in the category $\tilde\cG$ of directed graphs described in \S \ref{GraphSec}, 
with the property that $W'=\varphi_*(W)$, where the pushforward of the weight matrix is defined as
$$ \varphi_*(W)(u',v')=\sum_{\substack{u\,: \varphi(u)=u' \\ v\,: \varphi(v)=v'}} W(u,v)\, . $$

\smallskip

If the graph $G$ has parallel edges, namely different edges $e\neq e'$ with the same source and target,
$s(e)=s(e')=v$ and $t(e)=t(e')=v'$ then the function $W(v,v')$ represents a sum of the weights 
of all the edges connecting $v$ and $v'$. 

\smallskip

The monoidal structure on $\text{IDAG}^o$ defined above induces a monoidal structure on $\text{WIDAG}^o$ with
$$ (G,W)\oplus (G',W')=(G\oplus G', W\oplus W')\, , $$
with $G\oplus G'$ defined as above and
with $W\oplus W'$ given by the following row-column description: 
\begin{equation}\label{WWsum}
(W\oplus W')((u,o'),(v,o'))=W(u,v) \  \text{ and }  \  (W\oplus W')((o,u'),(o,v'))=W'(u',v')\,  . 
\end{equation}

\smallskip

The unit for this monoidal structure is
$$ (0,W_0)=((\emptyset, \emptyset, \{ o \}, \emptyset),W_0(o,o)=0)\, .$$ 

\smallskip

Note that, for an object $(G,W)\in \text{WIDAG}^o$, the condition that there exist a morphism $\varphi: (G,W) \to (0,W_0)$
amounts to requiring that the map that collapses the whole graph $G$ to the single output vertex $o$ satisfies
$\varphi_*(W)(o,o)=W_0(o,o)=0$, which means
$$  \sum_{(u,v)\in V\times V} W(u,v) =0 \, . $$
This condition can be seen as an inhibitory-excitatory balance condition over the whole network $G$.

\smallskip

On the other hand, for an object $(G,W)\in \text{WIDAG}^o$, the condition that there exist a morphism $\psi: (0,W_0) \to (G,W)$
implies that the inclusion of the output vertex in $G$ satisfies $\psi_*(W_0)(u,v)=W(u,v)$ for all $(u,v)\in V\times V$.
This implies that $W: V\times V \to \R$ is the trivial map that is identically zero on all of $V\times V$. Thus, the only
objects of $\text{WIDAG}^o$ with a morphism from $0$ are those with trivial weights. This shows that the unit $(0,W_0)$
of the symmetric monoidal structure is neither an initial nor a final object for $\text{WIDAG}^o$. 

\smallskip

The operad structure on $\text{IDAG}^o$ induces an operad structure on $\text{WIDAG}^o$, where the
composition operations $\circ_j$ that match the output vertex $o$ of $(G,W)\in \cO(n)$ to the $j$-th input
vertex $v'_j$ of $(G',W')\in \cO(m)$ give an object $(G\circ_j G', W\circ_j W')$, where $G\circ_j G'$ is the
composition in $\text{IDAG}^o$ and $W\circ_j W'$ with entries
\begin{equation}\label{WWcomp}
\begin{array}{ll} 
(W\circ_j W')(u,v)=W(u,v) & \text{when }  u,v\in V, \\[3mm]
(W\circ_j W')(u',v')=W'(u',v') & \text{when } u',v'\in V', \\[3mm]
(W\circ_j W')(u,v'_j)=W(u,o) & \\[3mm] 
(W\circ_j W')(o,v')=W'(v'_j,v') & 
\end{array}
\end{equation}
and zero otherwise. 
This composition rule for weights is 
compatible with the operad axioms. 
Both \eqref{WWsum} and \eqref{WWcomp} can be seen as obtained
by taking a block sum of the two matrices $W$ and $W'$ and then
identifying the indices of the vertices that are glued together, merging
the corresponding rows and columns.

\smallskip
\subsubsection{Category of DNNs}

The construction of the category $\text{WIDAG}^o$ provides a setting where we can describe  
computational resources based on DNNs and their compositions where the output of one
feeds into input of another, or where two independent ones feed to a single output
that integrates their separate inputs. These two types of serial and parallel composition are obtained, 
respectively, through the operadic composition laws of $\text{WIDAG}^o$ and the
monoidal structure. 

\smallskip

We interpret objects $(G,W)$ of $\text{WIDAG}^o$ as feed-forward neural networks, 
with vertices in $I$ representing the input layer, the unique
output $o$ representing the output layer, and the set $H$ of hidden nodes representing the
intermediate layers, with $W$ an initialization of weights. Note that 
we are allowing edges that skip layers.
It is known that every directed acyclic graph can be used as architecture for a 
feed-forward neural network in this way, when connections that skip layers are allowed.

\smallskip
\subsection{Threshold dynamics with leak term and the category of DNNs}\label{LeakSec}

We first look at the behavior of equation \eqref{catHopf1} (the first case with leak term) 
when the category of resources is taken to be $\cC=\text{WIDAG}^o$.

\smallskip

The category of network summing functors that we consider for our example of
categorical Hopfield dynamics is the category $\Sigma_{\text{WIDAG}^o}(\fG)$ for a given network
$\fG$. As we mentioned in \S \ref{IntroSec} (see \cite{ManMar} for a more detailed account),
for a given category of resources $\cC$ with a properad structure the
category $\Sigma_\cC(\fG)$ of network summing functors (also denoted by 
$\Sigma_\cC^{prop}(\fG)$ in \cite{ManMar}) has objects given by functors
$\Phi: P(V(\fG))\to \cC$, with $P(V(\fG))$ the category of subsets of $V(\fG)$ with
inclusions as morphisms, with the properties:
\begin{itemize}
\item summing property: $\Phi(\fG_1\sqcup \fG_2)=\Phi(\fG_1)\oplus \Phi(\fG_2)$ with $\Phi(\emptyset)=0$;
\item connectivity: for all $\fG'\in P(\fG)$
$$ \Phi(\fG') \in {\rm Obj}(\cC(\deg^{in}(\fG'), \deg^{out}(\fG'))\, , $$
with $\deg^{in/out}$ the in/out-degree, namely the number of oriented edges entering/leaving the
subgraph $\fG'$ from/to other vertices of $V(\fG)\smallsetminus V(\fG')$.
\item compositionality: for all $\fG', \fG' \in P(\fG)$, 
\begin{equation}\label{graftingPhi}
\Phi(\fG'\star \fG'') = \Phi(\fG') \circ_{E(\fG',\fG'')} \Phi(\fG''), 
\end{equation}
where $E(\fG',\fG'')\subset E(\fG)$ is the set of edges with one endpoint in $V(\fG')$ and the other in 
$V(\fG'')$ and $\circ_{E(\fG',\fG'')}$ is the properad composition that composes outputs to inputs along
these edges.
\end{itemize}

Here we want to consider the case where $\cC=\text{WIDAG}^o$. Since on 
$\text{WIDAG}^o$ we have an operad structure rather than a more general properad,
in order to have the correct compositional structure \eqref{graftingPhi} for 
these summing functors, we
will assume that the directed graphs $\fG$ that we consider here have, at each node, a single
output direction and an arbitrary number of inputs. 

\smallskip

We think of the network $\fG$ as describing the wiring of a population of neurons
(labelled by the vertices of $\fG$) with their synaptic connections. On the other hand,
at each vertex $v$, the object $(G_v,W_v)$ describes the DNN that models the
neuron labelled by $v$.

\smallskip

Note that here it is important to distinguish between the network $\fG$ over which the dynamics takes place,
and the networks describing the internal computational architectures assigned to the nodes of $\fG$, namely
the collection of objects $$\{ (G_v,W_v)=\Phi(v) \}_{v\in V(G)} \in {\rm Obj}(\text{WIDAG}^o) $$ that specify a compositional network summing functor $\Phi \in \Sigma_{\text{WIDAG}^o}(\fG)$.  We use the notation
$\fG$ for the underlying network (rather than $G$ as in \S \ref{IntroSec} above) to avoid any possible confusion. 

\smallskip 

We only assume here that the graph $\fG$ is a finite directed graph. We do not
require that it is acyclic. In general
we allow for the possibility that $\fG$ has multiple edges and looping edges. We also
do not require any particular structure of inputs and outputs for $\fG$.
On the other hand the graphs $G_v$ are in the category $\text{IDAG}^o$ hence
we assume they are acyclic, with a single output vertex, and with single
outgoing edges at other vertices, and with a single incoming flag at input vertices
and a single outgoing flag at the output vertex. 

\smallskip

\begin{figure}
 \begin{center}
 \includegraphics[scale=0.3]{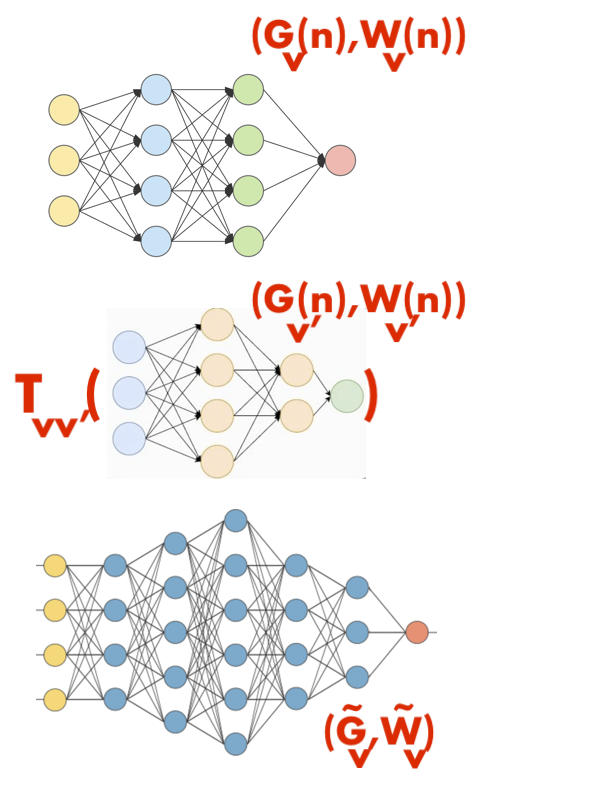}  \ \ \  \ \  \
 \includegraphics[scale=0.3]{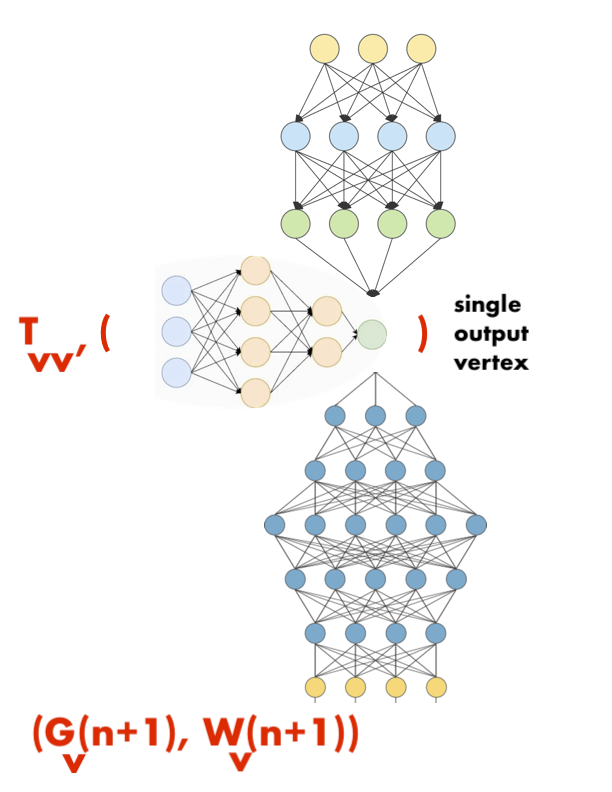}  
 \end{center} 
 \caption{The update rule \eqref{Hopf1GW} of the dynamics takes the objects  $(G_v(n),W_v(n))$
 and $(\tilde G_v, \tilde W_v)$ at the vertex $v$ and the 
 $T_{vv'}(G_{v'}(n),W_{v'}(n))$ and glues their outputs together to a single output
 to obtain the updated $(G_v(n+1),W_v(n+1))$. \label{ImageTryFig}}
 \end{figure}

We can then write the categorical Hopfield equations in the category of compositional
summing functors $\Sigma_{\text{WIDAG}^o}(\fG)$ as
\begin{equation}\label{Hopf1GW}
 (G_v(n+1),W_v(n+1))=(G_v(n),W_v(n)) \oplus \left( \oplus_{v'\in V(\fG)} T_{vv'} (G_{v'}(n),W_{v'}(n)) \oplus (\tilde G_v,\tilde W_v) \right)_+ \, .
\end{equation} 
Thus, here the next step of the dynamics takes the graph with weights $(G_v(n),W_v(n))$ of the current
step and glues it at the output with another graph, which is either the trivial single output graph when
the last term is below threshold and is a sum of the graphs with weights $(G_{v'}(n),W_{v'}(n))$, transformed
via the functors $T_{vv'}$ and the assigned (bias term) graph with weights $(\tilde G_v,\tilde W_v)$, when
this sum term is above threshold. This is illustrated in Figure~\ref{ImageTryFig}.

\smallskip

Here $(\tilde G_v,\tilde W_v)$ is an assigned summing functor which plays the role of a ``bias term" at each node $v\in V(\fG)$. This can be
taken to be $(0,W_0)$ at all $v$, for example. A possible choice of the collection $T=\{ T_{vv'} \}_{v,v'\in V(\fG)}$ of endofunctors of $\text{WIDAG}^o$ will
be discussed in \S \ref{EndoFuncSec} below. The threshold $(\cdot )_+$ imposes an inhibitory-excitatory balance condition on the
resulting DNNs 
\begin{equation}\label{bGvbWv}
(\bG_v,\bW_v)(n) :=\oplus_{v'\in V(\fG)} T_{vv'} (G_{v'}(n),W_{v'}(n)) \oplus (\tilde G_v,\tilde W_v).
\end{equation}
 Indeed the threshold detects whether there
are non-trivial morphisms, $${\rm Mor}_{\text{WIDAG}^o}((\bG_v,\bW_v)(n), (0,W_0))\neq \emptyset .$$ As discussed above this condition is satisfied provided that
$\bW_v$ satisfies the inhibitory-excitatory balance condition 
$$\sum_{a,b \in V(\bG_v)} \bW_v(a,b)=0. $$

\smallskip

When the balance condition above fails, the threshold replaces $(\bG_v,\bW_v)(n)$ with the unit $(0,W_0)$ so that the equation stabilizes, with
$$(G_v(n+1),W_v(n+1))=(G_v(n),W_v(n)). $$ 
Thus, the dynamics reaches a fixed point at the first $n\in \N$ for which the balance condition of $(\bG_v,\bW_v)(n)$ is violated. 

\smallskip

If the balance condition is not violated at any $n\in \N$, then a solution is given by a non-stationary sequence $(G_v(n),W_v(n))$. 
This can be seen as a sequence
$\Phi_n$ of compositional network summing functors in $\Sigma_{\text{WIDAG}^o}(\fG)$. One can ask if and in what sense this
solution may approach a fixed point.

\smallskip

As discussed in \cite{ManMar},
the evolution $\Phi_n \mapsto \Phi_{n+1}$ is implemented by an endofunctor $\cT$ of the category $\Sigma_{\text{WIDAG}^o}(\fG)$
that maps $\cT: \Phi \mapsto (T(\Phi)\oplus \tilde\Phi)_+$ where $\tilde\Phi$ is the summing functor defined by the local data
$(\tilde G_v,\tilde W_v)$, $T$ is the endofunctor implemented on the local data by the $T_{vv'}$, and the threshold is seen
as in \cite{ManMar} as an endofunctor (not necessarily monoidal) of $\hat\cC$. Thus, stationary solutions can be viewed as fixed points
of the endofunctor $\cT$. As we discussed in \S \ref{FixedPtSec} above, according to the general notion of
fixed points of endofunctors, this means here that stationary solutions are 
summing functors $\Phi\in \Sigma_{\text{WIDAG}^o}(\fG)$ that are isomorphic to
their image under this endofunctor, $\Phi \simeq \cT(\Phi)$. This condition corresponds to the existence of isomorphisms in $\cC$
$$ \eta_v: (G_v,W_v) \stackrel{\simeq}{\rightarrow} (G_v,W_v) \oplus \left( \bigoplus_{v'} T_{vv'} (G_{v'}, W_{v'}) \oplus (\tilde G_v, \tilde W_v) \right)_+\, , $$
where $\Phi_v=(G_v,W_v)$. 

\smallskip

Even though we have not specified so far a choice of the endofunctors $T_{vv'}$ of $\cC$, we can already make some general
comments about the behavior of solutions in this specific example of categorical Hopfield dynamics. Indeed, the fixed point
condition above, in the case where the balance condition holds on the right-hand-side, so that the threshold is the identity, in particular implies 
that the underlying directed graphs are related by an isomorphism 
$$ \eta_v: G_v \stackrel{\simeq}{\rightarrow} G_v \oplus \bG_v $$
with $(\bG_v,\bW_v)$ as above. By our definition of the monoidal structure in $\text{WIDAG}^o$, the graph
$G_v \oplus \bG_v$ has vertex set $V(G_v \oplus \bG_v)=V(G_v)\times \{ o' \} \sqcup \{ o \} \times V(\bG_v)$, where $o$ is the single
output vertex of $G_v$ and $o'$ is the single output vertex of $\bG_v$. The existence of an isomorphism between $G_v$ and $G_v\oplus \bG_v$
would require $V(G_v)\simeq V(G_v \oplus \bG_v)$, which can happen only if $\bG_V=\{ o' \}$ so that necessarily $(\bG_v,\bW_v)=(0,W_0)$.
Thus, we see that in this case the only fixed points are indeed those $(G_v,W_v)$ such that $(\bG_v,\bW_v)=(0,W_0)$. Note that, because $(\bG_v,\bW_v)$ is a monoidal sum
$$ (\bG_v,\bW_v)= \bigoplus_{v'} T_{vv'} (G_{v'}, W_{v'}) \oplus (\tilde G_v, \tilde W_v) \, , $$
in order to have $(\bG_v,\bW_v)=(0,W_0)$ we would need each term in the sum to be
$T_{vv'} (G_{v'}, W_{v'})=(0,W_0)$ and $(\tilde G_v, \tilde W_v)$. However, we know this
is not the case, either because the bias term $(\tilde G_v, \tilde W_v)$ is taken to be nontrivial,
or because at least some of the $T_{vv'} (G_{v'}, W_{v'})$ are nontrivial. We then reach the
conclusion that $(\bG_v,\bW_v)=(0,W_0)$ cannot be satisfied if $(\bG_v,\bW_v)$ satisfies the
balanced condition and no fixed points can occur in this case, since we cannot then have an
isomorphism between $(G_v,W_v)$ and $(G_v,W_v)\oplus (\bG_v,\bW_v)$. Thus, fixed
points only occur when $(\bG_v,\bW_v)$ violates the balanced condition. 

\smallskip

We can also consider the notion of strict fixed point recalled in \S \ref{FixedPtSec}. 
The summing functors $\Phi_n$ determined by the objects $\{ (G_v(n),W_v(n)) \}$ in turn determine 
a sequence of
vertices in the simplicial set given by the nerve $\cN(\Sigma_\cC(\fG))$. The endofunctor $\cT$ induces a map
of simplicial sets $\cT_\cN: \cN(\Sigma_\cC(\fG)) \to \cN(\Sigma_\cC(\fG))$. A fixed point 
$x=\cT_\cN(x)$ in  $\cN(\Sigma_\cC(\fG))$
corresponds to a strict fixed point for the endofunctor $\cT$, namely an object $\Phi\in \Sigma_\cC(\fG)$ such that
$\Phi =\cT(\Phi)$ (actually equal rather than just isomorphic, see \cite{Luzh}). 

\smallskip

On the other hand, since morphisms
in the category $\Sigma_\cC(\fG)$ are just invertible natural transformations, a vertex $x$ of  $\cN(\Sigma_\cC(\fG))$ such
that $x$ and $\cT_\cN(x)$ are connected by an edge of $\cN(\Sigma_\cC(\fG))$ correspond to 
a fixed point in the more
general sense considered above. Thus, we can take as a possible definition of an 
orbit $\{ \Phi_n \}_{n\in \N}$, with
$\Phi_{n+1}=\cT(\Phi_n)$, converging to a fixed point, the requirement that the induced 
sequence of vertices $x_n$
in $\cN(\Sigma_\cC(\fG))$ approaches either a single vertex or a sequence of vertices 
successively connected by an edge. 
Here we can use the topology of the realization $|\cN(\Sigma_\cC(\fG))|$ to measure proximity.  

 \smallskip
 
Observe then that, for the same reason discussed above, if we have a non-stationary sequence $(G_v(n),W_v(n))$, where the balanced
condition is satisfied at all $n$, we necessarily have, at each step
$$ \# V(G_v(n+1)) > \# V(G_v(n))\, , $$
so the sequence $\# V(G_v(n))$ is divergent at every $v\in V(\fG)$ where the balanced condition holds for all $n$.
This exclude the possibility that the corresponding sequence of vertices in $\cN(\Sigma_\cC(\fG))$ would stabilize to
either $x_n=\cT_\cN(x_n)$ or to $x_n$ and $\cT_\cN(x_n)$ connected by an edge (an isomorphism in $\Sigma_\cC(\fG)$) 
as both conditions would require the number of vertices of $G_v(n)$ to stabilize. 

\smallskip

Thus, regardless of the specific choice of the family of endomorphisms $T_{vv'}$, we see that in the case of the
first equation \eqref{catHopf1}, solutions converge to a fixed point (in finitely many steps) if and only if the
balanced condition on $(\bG_v, \bW_v)$ is violated after finitely many steps and the other solutions do not
converge to fixed points.

\smallskip

Assuming that $(G_v(n),W_v(n))$ satisfies the balanced condition, the violation of the balanced condition for 
$(\bG_v, \bW_v)$ depends on the choice of $T_{vv'}$ and of $(\tilde G_v, \tilde W_v)$. Since these are chosen
with the equation, we can just assume for simplicity that the objects $(\tilde G_v, \tilde W_v)$ do satisfy the
balanced condition, so that the possible violations are only coming from the choice of the endofunctors $T_{vv'}$
and not from the combined effect of the two choices. Thus, the question of the violation of the balanced condition
becomes a question of when the object 
$\oplus_{v'\in V(\fG)} T_{vv'} (G_{v'},W_{v'})$ fails to satisfy the balanced condition.

\smallskip
\subsection{Threshold dynamics without leak term and the category of DNNs}\label{NoLeakSec}
 
We now consider the second case (without leak term) of the categorical Hopfield dynamics given by equation \eqref{catHopf2} with $\cC=\text{WIDAG}^o$. 
We are assuming the same hypothesis about the underlying network $\fG$  and about the graphs $G_v$
as in \S \ref{LeakSec}.

\smallskip

In this setting without leak term, instead of \eqref{Hopf1GW} the equation we consider is of the form
\begin{equation}\label{Hopf2GW}
 (G_v(n+1),W_v(n+1))= \left( \oplus_{v'\in V(\fG)} T_{vv'} (G_{v'}(n),W_{v'}(n)) \oplus (\tilde G_v,\tilde W_v) \right)_+ \, .
\end{equation} 
Stationary solutions occur in two cases:
\begin{enumerate}
\item for pairs $(G_v,W_v)$ where the right-hand-side 
$\oplus_{v'\in V(\fG)} T_{vv'} (G_{v'},W_{v'}) \oplus (\tilde G_v,\tilde W_v)$
satisfies the balanced condition and
\begin{equation}\label{Hopf2fix}
 (G_v,W_v)= \oplus_{v'\in V(\fG)} T_{vv'} (G_{v'},W_{v'}) \oplus (\tilde G_v,\tilde W_v), 
\end{equation} 
\item  in cases where the right-hand-side does not satisfy the balanced condition and the datum
$(\tilde G_v,\tilde W_v)$ also does not satisfy the balanced condition. 
\end{enumerate}

\smallskip

In the second case the stationary solution is the unit $(0,W_0)$ of the monoidal structure. 
We focus on some particular examples of the first case, where depending on the
choice of the endofunctors $T_{vv'}$ and the datum $(\tilde G_v,\tilde W_v)$ we
find different behaviors of stationary solutions. In particular, in \S \ref{DecoupleSec}
we analyze a very simplified toy model case.

\smallskip

We use the same
notation $(\bG_v,\bW_v)$ as in \eqref{bGvbWv}. In the first case,
the fixed point condition \eqref{Hopf2fix} in particular implies that
$G_v \simeq \bG_v$ are isomorphic as directed graphs. 

\medskip
\subsubsection{Decoupled dynamics} \label{DecoupleSec}

In order to give a very basic example of the dynamics \eqref{Hopf2GW},
we make a drastic simplifying assumption on the form of the endofunctors $T_{vv'}$
that completely decouples the dynamics at the individual nodes $v\in V(\fG)$ from the
structure of the graph $\fG$. This is, of course, a very non-realistic assumption,
as in general one wants the $T_{vv'}$ to reflect the connectivity structure of $\fG$,
but it is useful for the purpose of presenting a simple explicit example. In particular,
this simplified setting of uncoupled dynamics will suffice to show (see \S \ref{EndoFuncSec}
below) that backpropagation in DNNs arises as a special case of our categorical Hopfield
dynamics. 

\smallskip

Let us then consider the case where the endofunctors $T_{vv'}$
have the form $T_{vv'} =T_v \delta_{v,v'}$ where here the Kronecker
delta $\delta_{v,v'}$ means that for $v\neq v'$ the endofunctor
$T_{vv'}$ maps all objects to the unit $(0,W_0)$. The endofunctors
$T_v$ are taken to be unital monoidal endofunctors of $\cC$. In particular
they map $(0,W_0)$ to itself. 
(If we want to think of the $T_{vv'}$ in terms of graph edges, this
would correspond to a network $\fG$ that has a looping edge at each vertex,
and this edge is solely responsible for the dynamics.)

\smallskip

We also assume that the endofunctors $T_v$ of $\cC$ act on objects
$(G_v,W_v)$ by maintaining the same underlying graph and updating
the weights,
\begin{equation}\label{TW}
T_v (G_v,W_v) = (G_v, T_v(W_v))\, . 
\end{equation}

\smallskip

Under these simplifying assumptions, we see by the same argument used
in the case of equation \eqref{Hopf1GW} that the isomorphism of directed
graphs $G_v \simeq \bG_v=G_v\oplus \tilde G_v$ is possible only if the
term $(\tilde G_v,\tilde W_v)=(0,W_0)$. We then have the fixed point
condition of the form
\begin{equation}\label{Tvfix}
(G_v,W_v) = (G_v, T_v(W_v))\, ,
\end{equation}
where (as discussed in \S \ref{IntroSec}) we interpret the ``$=$" sign
in the equations as isomorphism rather than equality, so that we
consider also non-strict fixed points (see \S \ref{FixedPtSec}).
An isomorphism $\varphi: (G_v,W_v)\to (G_v, T_v(W_v))$ is
an automorphism $\varphi: G_v \to G_v$, namely bijections 
$\varphi_V: V(G_v)\to V(G_v)$ and $\varphi_E: E(G_v)\to E(G_v)$
compatible with source, target, and $\epsilon$ morphisms (see
\S \ref{GraphSec}) and such that for all vertices $a,b\in V(G_v)$
$$ T_v(W_v)(\varphi(a),\varphi(b)) =  W_v(a,b) \, . $$
In particular, in \S \ref{EndoFuncSec} we focus on the case
of strict fixed points where $\varphi$ is the identity and
$T_v(W_v)=W_v$. 
This is a fixed point condition for the updating of the weights $W_v$
through the map $T_v$. 

\smallskip

Thus, the kind of dynamics we are considering here consists of performing
an update on the weights of the individual DNN machines $(G_v,W_v)$ 
located at each node $v$ of the network $\fG$, where the update of the weights 
is performed by the functor $T_v$.
For DDNs a typical example of update of the weights is the 
backpropagation mechanism, realized via a gradient descent algorithm.  
We will show in \S \ref{EndoFuncSec} that indeed backpropagation 
fits in the functorial formulation we are presenting here. 

\smallskip

Of course this case is oversimplistic as it assumes that the
updating of the weights of each machine $(G_v,W_v)$ is independent
of the inputs/outputs it received/transmits to the machines placed at
other nodes of $\fG$. However, it is useful to consider as a simplest example.

\smallskip
\subsubsection{Functorial gradient descent}\label{EndoFuncSec}

Let us consider a simple mechanism for updating the weights $W_v$ of
the DNN $(G_v,W_v)$ in terms of gradient descent with respect to
a given cost function. We write $F_v$ for the given relevant cost
function, which we assume is specific to the DNN $(G_v,W_v)$ and
therefore varies with the choice of $v\in V(\fG)$. We also fix a scale
$\epsilon >0$, the learning rate. This can be taken uniform over $V(\fG)$,
since this set is finite. Then the updating of the weights by gradient
descent take on the usual form
\begin{equation}\label{nablaW}
T_v(W_v) := W_v - \epsilon  \nabla_{W_v} F_v \, ,
\end{equation}
so that the condition that $W_v$ is a fixed point, $T(W_v)=W_v$, 
corresponds to the condition that the weights $W_v$ are a critical
point of the cost function, which under additional conditions on the
shape of the cost function can be assumed to be a minimum.

\smallskip

In our setting, we need to ensure that \eqref{nablaW} 
indeed defines an endofunctor of $\text{WIDAG}^o$. 
Given a cost function $F$ we map an object $(G,W)$ of $\text{WIDAG}^o$ to
$$ T(G,W)=(G, W-\epsilon \nabla_W F)\, . $$
Given a morphism $\varphi\in {\rm Mor}_{\text{WIDAG}^o}((G,W),(G',W'))$,
with $\varphi: G \to G'$ a morphism of directed graphs and
$W'=\varphi_*(W)$, we define $T(\varphi)$ as the same morphism
$\varphi: G \to G'$, since $T$ does not change the underlying graph. 
Thus, for this to give a morphism $T(\varphi)\in {\rm Mor}_{\text{WIDAG}^o}((G,T(W)),(G',T(W')))$
we just need to check the consistency
$$ T(W')= \varphi_* \, T(W)\, . $$
The right-hand-side is given by
$$ (\varphi_* \, T(W)) (u',v') = \sum_{\substack{u\,: \varphi(u)=u' \\ v\,: \varphi(v)=v'}} T(W) (u,v) 
 = \sum_{u,v} W(u,v) -\epsilon \sum_{u,v} \nabla_{W(u,v)} F $$ 
 which agrees with $W' (u',v') -\epsilon \nabla_{W'(u',v')} F = T(W') (u',v')$. 

\smallskip

Thus, with this particular choice of the endofunctors $T_{vv'}=T_v \delta_{v,v'}$,
the stationary solutions of the equation \eqref{Hopf2GW} are given by
$\{ (G_v,W_v) \}$ where the graphs $G_v$ are as in the chosen initial
datum of the equation, $G_v(n)=G_v(0)$, for all $n\in \N$, and where 
the weights $W_v$ are minima of assigned cost functions $F_v$ (we think
of the cost functions as being part of the data specifying the functor $T_v$).
In this case we have a notion of convergence of a solution to a fixed point,
by the requirement that the sequence of weights $W_v(n)$ converge to a 
minimum of $F_v$. 

\smallskip

Note that the type of functoriality we are considering here is different from
the functorial backpropagation discussed in \cite{FoSpiTu}. The latter is
more closely related to our operadic compositional structure than to the
functoriality with respect to the category $\cC=\text{WIDAG}^o$ discussed here.

\bigskip
\bigskip
\subsection*{Acknowledgment} The author is partially supported by
NSF grants DMS-1707882 and DMS-2104330
and by FQXi grants FQXi-RFP-1 804 and FQXi-RFP-CPW-2014, 
SVCF grant 2020-224047. 


\end{document}